\documentclass[prb,showpacs,floatfix,twocolumn,superscriptaddress]{revtex4}
\usepackage{graphicx}
\usepackage{dcolumn}
\begin{document}

\newcommand*{\YBCO}{YBa$_2$Cu$_3$O$_{7-\delta}$\/}
\newcommand*{\LCCO}{La$_{1.9}$Ca$_{1.1}$Cu$_2$O$_{6+\delta}$\/}
\newcommand*{\LSCO}{La$_{2-x}$Sr$_x$CuO$_{4+\delta}$\/}
\newcommand*{\LSCCO}{La$_{1.85}$Sr$_{0.15}$CaCu$_2$O$_{6+\delta}$\/}
\newcommand*{\LSxCCO}{La$_{2-x}$(Ca,Sr)$_x$CaCu$_2$O$_{6+\delta}$\/}
\newcommand*{\BSCCO}{Bi$_2$Sr$_2$CaCu$_2$O$_{8+\delta}$\/}
\newcommand*{\BSCO}{Bi$_2$Sr$_2$CuO$_{6+\delta}$\/}
%
%
\title{Infrared properties of \LSxCCO\ single
crystals}
%
%

\author{N. L. Wang}
\email{nlwang@aphy.iphy.ac.cn}%
\author{P. Zheng}
\author{T. Feng}
\affiliation{Institute of Physics, Chinese Academy of Sciences,
P.~O.~Box 603, Beijing 100080, P.~R.~China}
\author{G. D. Gu}
\author{C. C. Homes}
\author{J. M. Tranquada}
\affiliation{Department of Physics, Brookhaven National
Laboratory, P.~O.~Box 5000, Upton, NY11973-5000, USA}
\author{B. D. Gaulin}
\author{T. Timusk}
\affiliation{Department of Physics and Astronomy, McMaster
University, Hamilton, Ontario L8S 4M1, Canada}
\date{November 8, 2002}
%
%
\begin{abstract}
The in-plane optical properties of two crystals of the bilayer
cuprate \LSxCCO, one with excess Ca and $x=0.10$ and the other with
Sr and $x=0.15$, were investigated over the frequency range of
45--25000~cm$^{-1}$. The optical conductivity has been derived from
Kramers-Kronig transformation. Each crystal exhibits a peak at around
15000~cm$^{-1}$ which corresponds to the charge-transfer gap of the
parent insulator.  With increasing carrier density, spectral weight
shifts from the CT excitation to the low-$\omega$ region.   For the
superconducting sample ($x=0.15$), the optical conductivity displays a
peak in the far-infrared region, which shifts toward zero frequency with
decreasing temperature. The temperature-dependent behavior favors a
dynamical localization picture. A ``pseudogap'' feature is observed in the
low-frequency reflectance and the scattering rate spectra. Both the
energy scale and the temperature dependence of the ``pseudogap'' are
similar to other bilayer cuprates.
\end{abstract}


\maketitle

%
%
A common structural feature in all high-$T_c$ cuprates is the
presence of CuO$_2$ planes.  High-temperature superconductivity
depends not only on the carrier densities in the CuO$_2$ planes,
but also on the number of CuO$_2$ planes in a unit cell ($n$).
Within each family of cuprates, the superconducting transition
temperature ($T_c$) increases with layer number for $n\leq3$.
This is well illustrated in Bi-, Tl-, and Hg-based systems.  On
the other hand, the transition temperatures for different families
differ considerably. Among the most studied systems, the transition
temperature $T_c$ for bilayer materials \YBCO\ and \BSCCO\ at
optimal doping exceeds 90~K, whereas the maximum $T_c$s for
single-layer materials \LSCO\ and \BSCO\ are around $30-40$~K.
However, not all single-layer cuprates have such low transition
temperatures.  The maximum $T_c$ for Tl$_2$Ba$_2$CuO$_{6+\delta}$
also reaches 90~K.  For this reason, we classify \LSCO\ and \BSCO\
as low-$T_c$ compounds within the high-$T_c$ cuprates.  At
present, the reason for different $T_c$s in those materials is
unclear.

Among all known bilayer materials, the La-based system \LCCO\
(La2126) phase could be regarded as the simplest one.\cite{Cava} The
structure consists of a pair of pyramidal Cu-O layers facing one
another, which are the only electronically active elements.
Therefore, it is expected that this material could provide
essential information about bilayer cuprates. Unfortunately, it
has proven to be very difficult to make the material
superconducting. The reported superconducting transition
temperatures for single crystals prepared under high oxygen
pressure, or with Sr substitution for La, are below 50~K
(Ref.~\onlinecite{Watanabe}).  For polycrystalline samples, $T_c$
can be a bit higher (around 60~K) than for the single
crystals,\cite{Cava} but it is not clear why higher $T_c$s are not
observed.  At present, La2126 must be considered as a low-$T_c$ case
among the bilayer high-$T_c$ cuprates. It is of interest to
investigate physical properties of this system and to compare them with
the single-layer \LSCO\ in the same family, as well as with other bilayer
cuprates.

Owing to the difficulty of obtaining superconducting compounds,
relatively few physical measurements have been made on this system. In
this work we present the in-plane reflectivity and optical conductivity
data for single crystals of \LCCO\ and \LSCCO . We shall show how
optical spectra evolve with temperature for the more-highly doped,
slightly superconducting sample.  To our knowledge, such
temperature-dependent work has not been done previously. The only
reported in-plane optical work\cite{Shibata} was carried out at room
temperature and for frequencies higher than 250~cm$^{-1}$.

Large single crystals of \LSxCCO\ were grown by the
traveling-solvent floating-zone technique.  The Ca-doped $x=0.10$
sample was grown in oxygen atmosphere at ambient pressure, while
the Sr-doped $x=0.15$ crystal was grown at an oxygen partial
pressure of 10~bar.  Based on the dopant concentration, and
ignoring the uncertainty in the oxygen stoichiometry, the hole
concentrations per CuO$_2$ plane are 0.05 and 0.075, respectively,
spanning the threshold concentration for superconductivity in
\LSCO.  Magnetization measurements with a superconducting quantum
interference device indicate that the $x=0.10$ sample is not
superconducting down to the lowest measurement temperature
$\sim\!2$~K, but that the Sr-substituted sample exhibits a weak
superconducting response with an onset temperature of 28~K.  For a
given value of $x$, Kinoshita and Yamada\cite{kino92a} found that
both $T_c$ and the Meissner signal increase with oxygen content
(controlled by annealing in an oxygen atmosphere of increasing
pressure).  Our $x=0.15$ crystal, grown in 10~bar O$_2$, appears
comparable to their sintered sample annealed in 50~atm (49~bar)
O$_2$.

The polarized reflectance measurements from 45 to 25000~cm$^{-1}$ for
${\bf E}\| a$-axis were carried out on a Bruker 66v/S
spectrometer on polished surfaces of crystals, which were mounted
on optically black cones in a cold-finger flow cryostat using an
\textit{in situ} overcoating technique.\cite{Homes}  The optical
conductivity spectra were derived from the Kramers-Kronig
transformation. Since the high-$\omega$ extrapolation affects the
conductivity spectra, especially the oscillator strength of the
charge-transfer (CT) excitation, we connect the reflectance
spectra to the high frequency data of \LSCO\ by Uchida {\it et
al.}\cite{Uchida}  The Hagen-Rubens relation was used for the low
frequency extrapolation. It is found that different choices of dc
conductivity values have minor effect on the conductivity in the
measured frequency region.

%
%
\begin{figure}[t]
\centerline{\includegraphics[width=3.2in]{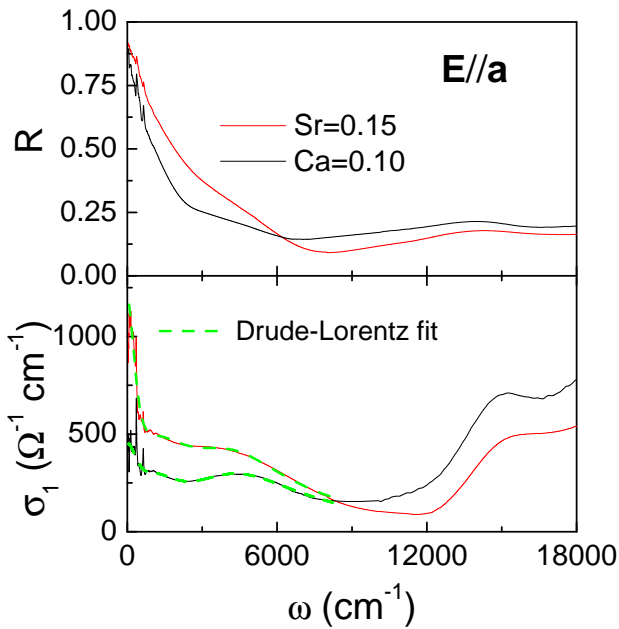}}%
\vspace*{-0.2cm}%
\caption{The frequency dependent reflectance and conductivity of
\LSxCCO\ with x=0.10 (Ca-doped) and x=0.15 (Sr-doped) at room
temperature. The dash curves are Drude-Lorentz
fits.}%
\label{fig1}
\end{figure}

Figure 1 shows the in-plane reflectance and optical conductivity
spectra of the two crystals at room temperature.  Below the
reflectance edge at about 8000~cm$^{-1}$, the reflectance of the
Ca-doped sample is substantially lower than that of the
Sr-substituted sample, as one would anticipate based on the different
nominal hole concentrations.  In accord with this, the optical
conductivity at low frequencies is considerably larger in the Sr-doped
sample. Detailed analysis of the low-$\omega$ response will be presented
in following paragraphs.  At higher frequencies, the conductivity has two
distinct absorption features: a Cu-O CT excitation near 15000~cm$^{-1}$
and mid-infrared bands.  With increasing carrier concentration,
spectral weight is shifted from the CT excitation to the low-$\omega$
region.  Comparisons with measurements on \LSCO\
(Ref.~\onlinecite{Uchida}), \YBCO\ (Ref.~\onlinecite{coop93}), and
Y-substituted \BSCCO\ (Ref.~\onlinecite{Wang}) indicate that the
conductivity spectra are consistent with samples near the
hole-concentration threshold for superconductivity.

The formation of a mid-infrared band is often seen in strongly
correlated electronic materials.  In doped Mott insulators, the
mid-infrared peak is the first feature to appear at low doping; with
increasing doping, weight grows at lower frequencies and eventually
a Drude-like peak centered at $\omega=0$ develops.\cite{Okimoto} This
kind of evolution has been well documented in the single-layer \LSCO\
cuprate.\cite{Uchida}  The observed change in bilayer \LCCO\ is
consistent with the general feature.  The accumulation of low-$\omega$
spectral weight could be quantitatively analyzed from the partial
sum-rule, $N_{eff}(\omega)= (2m_eV/\pi
e^2)\int_0^\omega\sigma_1(\omega^\prime)d\omega^\prime$, where
$m_e$ is the bare electron mass and $V$ is the volume of the unit
cell.  Usually, an integral of the spectral weight below the
frequency of reflectance or conductivity minimum would give
approximately the overall plasma frequency. However, since the
$\sigma_1(\omega)$ contains obviously different components, such
integration may overcount the contributions of free carriers.  A
Drude-Lorentz analysis would be more appropriate in this case.
The general formula for the optical conductivity of the
Drude-Lorentz model is\cite{Timusk}
\begin{equation}
   \sigma_1(\omega)=
   {\omega_p^{*2}\over4\pi}{\Gamma_D\over \omega^2+\Gamma_D^2} +
   \sum_j{\omega_{p,j}^2\over4\pi}
   {\Gamma_j\omega^2\over(\omega_j^2-\omega^2)^2+\omega^2\Gamma_j^2}.%
\label{chik}
\end{equation}
where $\omega_p^*$ and $\Gamma_D$ in the Drude term are the plasma
frequency and the relaxation rate of the free charge carriers,
while $\omega_j$, $\Gamma_j$, and $\omega_{p,j}$ are the resonance
frequency, the damping, and the mode strength of the Lorentz
oscillators, respectively.  As shown in Fig.~1, the main features
in $\sigma_1(\omega)$ below the frequency of the reflectance
minimum can be well reproduced by the combination of one Drude
component and two Lorentz oscillators.  The fitting parameters in
the Drude-Lorentz model for the two samples are shown in Table~I.
The overall evolution of optical spectra with doping is similar to
other high-$T_c$ cuprates.  In the following we shall mainly focus
on the temperature-dependent behavior of the superconducting
sample in the low-energy region.

%
%
\begin{table}
\caption{The parameters of the Drude-Lorentz fit to the room temperature
data for the two samples. (All quantities are in units of cm$^{-1}$.) }
\begin{ruledtabular}
\begin{tabular}{lcccccccc}
  Sample&$\omega_p^*$&$\Gamma_D$&$\omega_1$&$\Gamma_1$&$\omega_{p,1}$ &
         $\omega_2$&$\Gamma_2$&$\omega_{p,2}$ \\
  \cline{1-1} \cline{2-3} \cline{4-6} \cline{7-9}
  Ca=0.10 & 4030 & 628 & 1400 & 2400 & 5100 & 4750 & 6000 & 9360 \\
  Sr=0.15 & 5575 & 433 & 1560 & 3140 & 8140 & 4600 & 5410 & 9230 \\
\end{tabular}
\end{ruledtabular}
\end{table}

Figure~2 shows the reflectance and conductivity spectra below
2000~cm$^{-1}$ at different temperatures for the Sr=0.15 sample.
In general, the temperature dependence of the optical spectra shows a
metallic response, with the low-$\omega$ conductivity becoming
significantly enhanced with decreasing temperature.  At the same time,
$\sigma_1(\omega)$ at 200~K and 300~K exhibits a peak at finite
frequency, reaching 150~cm$^{-1}$ at 300~K, in contrast to the $\omega =
0$ peak expected for standard Drude behavior.

%
%
\begin{figure}[t]
\centerline{\includegraphics[width=3.2in]{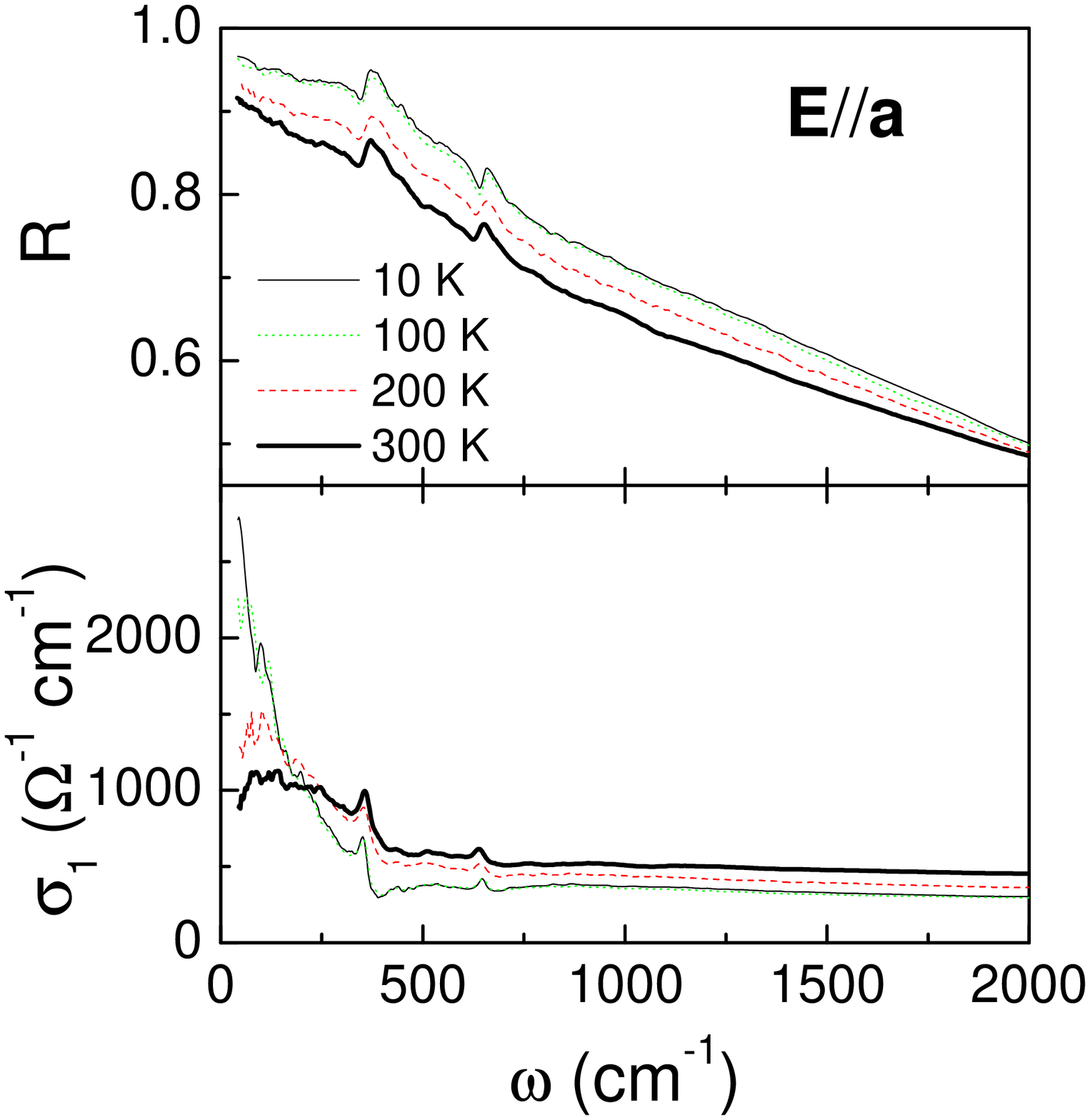}}%
\vspace*{-0.2cm}%
\caption{The frequency dependent reflectance and conductivity of
\LSCCO\ at different temperatures.}%
\label{fig2}
\end{figure}

A peak at finite frequency has been observed in a number of
high-$T_c$ cuprates\cite{Tsvetkov,Timusk2,Basov,
Basov2,McGuire,Takenaka} as well as in many other correlated
materials.\cite{Takenaka2,Kostic}  Several possibilities have been
suggested for the origin of the peak.  The simplest one is that it
is related to carrier localization caused by disorder, defects, or
impurities, i.e. Anderson localization.\cite{Timusk2,Basov,Basov2}
This interpretation received support from several studies where a
finite-frequency peak appeared when defects or impurities were
deliberately introduced into samples.\cite{Basov,Basov2,Wang2} For
example, for the pure \YBCO\ at optimal doping, there is no such
peak at finite energy; however, when some impurities such as Zn
were introduced into the sample, or when the sample was irradiated
by high-energy ions, the peak appeared at $\omega \neq 0$.  This
picture cannot explain those cases where there are no apparent
defects or impurities\cite{Kostic}; furthermore,  for
Anderson-type localization the effect should become more
significant at low temperature, while in many experiments,
including the present study, the peak was observed only at high
temperatures.  Nevertheless, the comparison to systems with
defects is relevant in our case, as diffraction studies of
\LSxCCO\ have found a tendency for 5 to 15\%\ of the Ca sites,
located between the CuO$_2$ bilayers, to be replaced by
La.\cite{izum89,saku91,kino92b,Ulrich}

A finite-energy peak has also been observed in systems with static charge
stripes, such as Nd-doped \LSCO\ (Refs.~\onlinecite{Tajima,Dumm}) and in
Nd-free \LSCO\ in which one might expect the existence of dynamical
stripe fluctuations.\cite{Venturini,Lucarelli}  Recent neutron
scattering measurements on \LSxCCO\ have found no evidence of any
charge ordering.\cite{Ulrich,Tranquada}  At the same time, elastic
scattering from antiferromagnetic clusters has been
observed,\cite{Ulrich,Tranquada} clearly indicating some form of
inhomogeneity.

Recently, Takenaka {\it et al.}\cite{Takenaka,Takenaka2} have
studied the finite-energy-peak phenomenon in \LSCO\ and
La$_{1-x}$Sr$_x$MnO$_3$.  They linked its occurrence to the
condition that the resistivity at high temperature exceeds the
Mott criterion $\rho_{Mott}$, which corresponds to the point at
which the quasiparticle mean free path $\ell$ becomes comparable
to the Fermi wavelength $\lambda_F$=$2\pi/k_F$
(Refs.~\onlinecite{Mott,emer95a}). Distinct from the low-$T$
Anderson localization caused by elastic scattering due to disorder
or impurities, the high-$T$ phenomenon is called ``dynamic''
localization, and is attributed to strong inelastic scattering.
Our present results appear to correspond to the latter case.

It is interesting to compare \LSCCO\ with other heavily underdoped
cuprates.  One way in which such comparisons are often made is to
plot the scattering rate vs.\ frequency, obtained by applying the
extended Drude model to the optical conductivity.\cite{Puchkov} In
optimally-doped cuprates, the scattering rate is linear in both
frequency and temperature above $T_c$ and develops a ``gap-like''
suppression below $T_c$. In underdoped samples, the scattering
rate depression sets in well above transition temperature.  The
depression of the scattering rate at $T>T_c$ has frequently been
associated with pseudogap phenomena\cite{Puchkov}; however, it
should be noted that no depression of the low-energy effective
density of in-plane charge carriers has been clearly demonstrated
in the pseudogap regime of any cuprates, contrary to the situation
for charge motion along the $c$-axis.\cite{home93}

%
%
\begin{figure}[t]
\centerline{\includegraphics[width=3.2in]{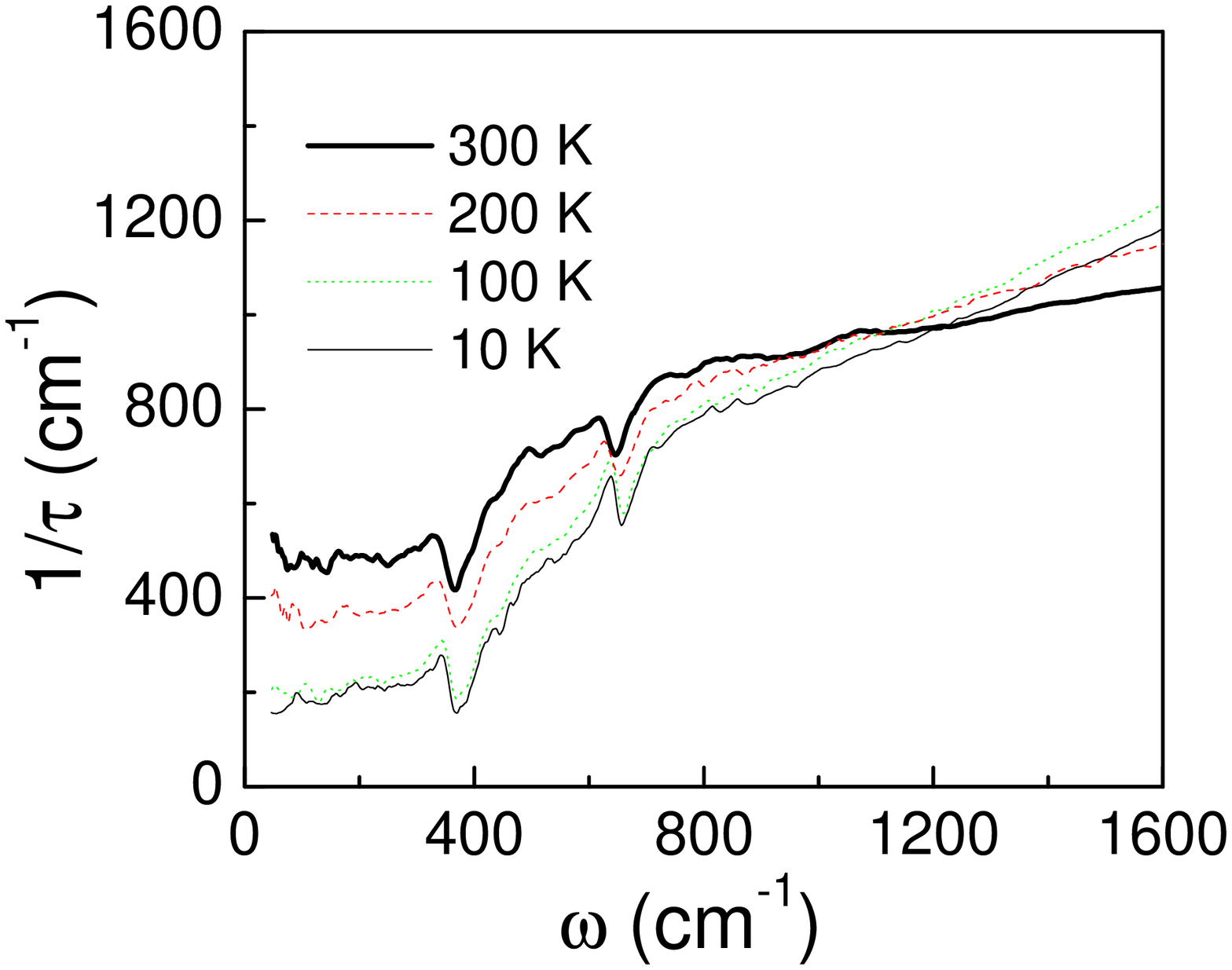}}%
\vspace*{-0.2cm}%
\caption{The frequency-dependent scattering rate of \LSCCO\ at
different
temperatures.}%
\label{fig3}
\end{figure}

Figure~3 shows the scattering rate spectra at different
temperatures for the \LSCCO\ sample being obtained from the
extended Drude model using the plasma frequency of 5575~cm$^{-1}$.
We can see that the scattering rate deviates from the
linear-$\omega$ dependence below 700~cm$^{-1}$ even at room
temperature, an energy scale very similar to that found in other
underdoped cuprates.\cite{Puchkov}  There is a substantial
residual scattering of about 480~cm$^{-1}$ at low frequencies
which is in good agreement with the 433~cm$^{-1}$ width of the
low-$\omega$ peak in $\sigma_1(\omega)$.  At lower temperatures,
the residual scattering decreases. In accord with this, we find
the narrowing of the low-$\omega$ peak in $\sigma_1(\omega)$.  In
the raw data of the reflectance, the ``gap-like'' feature of the
relaxation rate appears as the shoulder-like structure, which can
also be seen clearly in Fig.~2.

Finally, we make some comments about the superconducting
condensate.  The difference between the reflectance at low
temperature and that at $T_c$ in low frequencies (within the gap
energy) could be taken as a rough measure of the superconducting
condensate density.\cite{Timusk3}  From the reflectance below and
above the superconducting transition temperature, no obvious
difference could be detected, indicating that the condensed
carrier density is very small in this sample. Within current
understanding, the superconducting gap reflects the pairing
strength, while the condensed carrier density is an indication of
the phase stiffness of the pairing.\cite{uemu89,emer95b,Feng} Our
result suggests that the strength of pairing coherence is quite
weak. This can only partially be ascribed to the relatively low
carrier density doped into the material. The low superconducting
condensate is qualitatively consistent with the low $T_c$ of the
material.

In summary, we have determined the in-plane optical properties of two
\LSxCCO\ crystals. Both exhibit a peak at around 15000~cm$^{-1}$ which
corresponds to the charge-transfer gap of the parent insulator.  With
increasing carrier density, spectral weight shifts from the CT
excitation to the low-$\omega$ region.  For the superconducting
sample, the optical conductivity displays a peak in the far-infrared
region that shifts towards zero frequency with decreasing temperature.
The feature is similar to underdoped \LSCO, suggesting some
dynamical localization.  The low-frequency reflectance has some
knee structure between $400 - 700$~cm$^{-1}$, apparent at all measured
temperatures; similar features in other cuprates have been interpreted as
evidence of a pseudogap state.
Furthermore, the superfluid density is very low, consistent with the low
$T_c$, suggesting that the strength of pairing coherence is quite weak.

This work was in part supported by National Science Foundation of
China (No.10025418). The work at Brookhaven National Laboratory
was supported by the U.S. Department of Energy under contract
DE-AC02-98CH10886.
%
%

\end{document}